\documentstyle[12pt,epsfig]{article}
\begin{document}
\baselineskip=23pt

\rightline{BIHEP-TH-2002-17}
\bigskip
\begin{center}
{\Large\bf Dynamics of Massive Scalar Fields in dS Space\\
  and the dS/CFT Correspondence}

\bigskip
\bigskip

Zhe Chang\footnote{Email: changz@mail.ihep.ac.cn.}  and
Cheng-Bo Guan\footnote{Email: guancb@mail.ihep.ac.cn.}

{\em
Institute of High Energy Physics, Academia Sinica\\
P.O.Box 918(4), Beijing 100039, China}

\end{center}

\bigskip
\bigskip

Global geometric properties of dS space are presented explicitly in various coordinates. 
A Robertson-Walker like metric is deduced, which is convenient to be used in study of 
dynamics in dS space.
Singularities of wavefunctions of massive scalar fields at boundary are
demonstrated. A bulk-boundary propagator is constructed by making use of the solutions 
of equations of motion.
The dS/CFT correspondence and the Strominger's mass bound is shown.

\newpage

\section{Introduction}

Recent astronomical observations of supernovae and cosmic microwave background
\cite{Hi1}-\cite{co3} indicate that the universe is accelerating and can be well 
approximated by a world with a positive cosmological constant. If the universe would 
accelerate indefinitely, the standard cosmology leads to an asymptotic de Sitter (dS) 
universe. De Sitter space expands so rapidly that inertial observers see an event horizon. 
Like the Bekenstein-Hawking entropy\cite{BEK} of a black hole, the dS entropy
\cite{FIS}-\cite{BAN} can be written as

$$S=\frac{A}{4G}~,$$
where $G$ is Newton's constant, and $A$ is area of the event horizon. The holographic
principle then implies that the Hilbert space of quantum gravity in dS is finite dimensional.
As pointed out by Witten\cite{WIT}, if the Hilbert space of quantum gravity really has a 
finite dimension this gives a strong hint that the general relativity cannot be quantized 
and must be derived from a more fundamental theory. However, persistent efforts by many 
researchers have so far all failed to find any clear-cut way to get dS from superstring 
theory or $M$-theory. In fact, there is no positive conserved energy and there cannot be 
unbroken supersymmetry in dS space. Thus, a universe with an positive cosmological constant
poses serious challenges for superstring theory and $M$-theory. However, string theory may 
not be the only route to understanding of dS space. In fact, the AdS/CFT correspondence 
was first discussed from a general analysis of the asymptotic symmetries of  AdS\cite{BH3}. 
One can suspect that the AdS/CFT correspondence is a manifestation of the more fundamental 
holographic principle. Along this direction, in the similar sense of the AdS/CFT duality, 
Strominger \cite{STR} has suggested dS/CFT correspondence to relates quantum gravity on 
dS space with boundary conformal field theory\cite{MIN}-\cite{VOL}.

In this paper, we study dynamics for massive scalar fields in dS by solving exactly the 
equations of motion. Global geometric properties of dS is presented in various coordinates. 
A Robertson-Walker like metric is deduced from the familiar $SO(1,n+1)$ invariant one in 
an $(n+2)$-dimensional embedding space. Singularities of wavefunctions at the boundary are 
demonstrated explicitly. A bulk-boundary propagator  is constructed explicitly by making 
use of the solutions of equations of motion for massive scalar fields. The dS/CFT 
correspondence and the Strominger's mass bound is shown.

The paper is organized as follows. In Section 2, global geometric properties of dS in various 
coordinates are presented. The Penrose diagram is drawn. Relations between the explicit 
$SO(1,n+1)$ invariant metric and the Hua's metric\cite{Hua,Lu} are shown. 
In Section 3, We solve exactly the equations of motion for massive scalar fields 
by making use of the variable-separating method\cite{LG}. 
A bulk-boundary propagator is constructed in section 4.  The dS/CFT correspondence and 
the Strominger's mass bound is discussed.

\section{Penrose diagram and global geometric properties}

In an $(n+2)$-dimensional embedding space, the $(n+1)$-dimensional dS space can be written as

 \begin{equation}\label{sphere}
  \xi^0\xi^0-\sum_{i=1}^n\xi^i\xi^i-\xi^{n+1}\xi^{n+1}=-1~,
 \end{equation}
with induced metric 
 \begin{equation}\label{metric}
 ds^2=d\xi^0d\xi^0 - \sum_{i=1}^{n}d\xi^id\xi^i -d\xi^{n+1}d\xi^{n+1}~.
 \end{equation}
 \newline
It is easy to check that the isometric symmetry of dS$_{n+1}$ is $SO(1,n+1)$.
It is not difficult to see that, at least, $(n+1)$ charts of coordinates ${\cal U}_\alpha$
($\xi^{\alpha}\not= 0$, $\alpha=1,~2,~\cdots,~n+1$) should be needed to describe 
dS$_{n+1}$ globally.
 \newline
In the chart dS$_{n+1}\cap {\cal U}_\alpha$, we introduce a coordinate

 \begin{equation}
  x^\mu=\frac{\xi^\mu}{\xi^{\alpha}}~,~~~~~~(\mu\not=\alpha;~~\xi^{\alpha}\not=0)~.
 \end{equation}
for both the parts ${\cal U}^+_\alpha(\xi^{\alpha}>0)$ and ${\cal U}^-_\alpha(\xi^{\alpha}<0)$~.
 \newline
The chart dS$_{n+1}\cap {\cal U}_\alpha$, in the coordinate $(x^\mu)$, is described by
 \begin{equation}
 \begin{array}{l}
 {\rm dS}_{n+1}\cap {\cal U}_\alpha:~~~~~~\sigma(x^\mu,x^\nu)>0~,\\[0.5cm]
 \sigma(x^\mu,x^\nu)\equiv 1+\displaystyle\sum_{\mu,\nu\not=\alpha}\eta_{\mu\nu}x^\mu x^\nu ~,
 \eta={\rm diag}(-1,~\underbrace{1,~1,~\cdots,~1}\limits_{n+1})~.
 \end{array}
 \end{equation}
 \newline
At the overlap region dS$_{n+1}\cap{\cal U}_\alpha\cap {\cal U}_\beta$ of the two
charts ${\cal U}_\alpha$ and ${\cal U}_\beta$, we have relations between the coordinates
$(x^\mu)\in {\cal U}_\alpha$ and $(y^\nu)\in {\cal U}_\beta$

 \begin{equation}
 y^\alpha=\frac{1}{x^\beta}~,~~~~~~y^\nu=\frac{x^\nu}{x^\beta}~~~(\nu\not=\alpha).
 \end{equation}
This shows clearly a differential structure of dS space.
 \newline
The boundary $\overline{M}^n$ of the slice dS$_{n+1}\cap{\cal U}^+_\alpha$ of dS is
 \begin{equation}
 \overline{M}^n:~~1+\eta_{\mu\nu}x^\mu x^\nu=0~.
 \end{equation}
 \newline
In the coordinate $(x^{\mu})$, the metric is reduced as

 \begin{equation}
 ds^2=-\displaystyle\frac{d{x}J(I-{x}'{x}J)^{-1}d{x}'} {1-{x}J{x}'}~,
 \end{equation}
where 
  $J\equiv{\rm diag}(-1,~\underbrace{1,~1,~\cdots,~1}\limits_{n})~$,
  $x\equiv (x^0,x^1,\cdots, x^{\alpha-1},x^{\alpha+1},\cdots,x^n)~$, 
and $x'$ the transport of the vector $x$. This is
obviously a generalization of the Hua's metric for manifolds with Lorentz signature.

It should be noticed that the transformations ${\cal D}$ among the coordinate variables
$(x^\mu,~\mu\not=0)$ form a group ${\cal D}\in SO(n)$ as subgroup of $SO(1,n+1)$.
Thus, we can refer to these transformations as space-like ones.
The $x^0$ element of $(x^\mu)$ is not invariant under the so-called 
space-like transformations. We can not got a time-like Killing vector
using of the $x^0$. To draw explicitly time-like geodesics, one should investigate
space-like transformation invariance variables. In fact, we find that
under the transformations ${\cal D}$,
$\xi^0(\equiv\sigma^{-\frac{1}{2}}(x^{\mu},x^{\nu})x^0)$ is invariant.
Therefore, it is convenient to use the coordinate $(\xi^0,~x^1,~\cdots,~x^n)$.

By making use of the relations between the coordinates $(x^\mu)$ and $(\xi^0,~x^i)$

 \begin{equation}
 \begin{array}{l}
   \displaystyle \sigma(x^\mu,x^\nu)=\frac{1+{\bf x}{\bf x}'}{1+\xi^0\xi^0}~,\\[0.5cm]
   \displaystyle x^0x^0=\frac{\xi^0\xi^0}{1+\xi^0\xi^0}(1+{\bf x}{\bf x}')~,
   \end{array}
 \end{equation}
 \newline
we get a deduced Robertson-Walker like metric in terms of the coordinate $(\xi^0,~x^i)$

 \begin{equation}\label{PRW1}
  ds^2=\frac{d\xi^0d\xi^0}{1+\xi^0\xi^0}-(1+\xi^0\xi^0)
       \frac{d{\bf x}(I+{\bf x}'{\bf x})^{-1}d{\bf x}'}{1+{\bf x}{\bf x}'}~,
 \end{equation}
where the vector ${\bf x}$ denotes $(x^1,~x^2,~\cdots,~x^{\alpha-1},~x^{\alpha+1},
~\cdots,~x^{n+1})$ and ${\bf x}'$ the transport of the vector ${\bf x}$. 
 \newline
In the spherical coordinate $(x^1,~x^2,~\cdots,~x^{\alpha-1},~x^{\alpha+1}, ~\cdots,~x^{n+1})
\longrightarrow (\rho,~\theta_1,~\cdots,~\theta_{n-1})$, the Robertson-Walker like
metric is of the form
 \begin{equation}
  ds^2=\frac{d\xi^0d\xi^0}{1+\xi^0\xi^0}
       -(1+\xi^0\xi^0)\left[(1+\rho^2)^{-2}d\rho^2
       +(1+\rho^2)^{-1}\rho^2d{\Omega^2_{(n-1)}}\right] ~,
 \end{equation}
where $d\Omega^2_{(n-1)}$ is the metric on $(n-1)$-sphere.
 \newline
The boundary $\sigma(x^{\mu},x^{\nu})=0$ of dS space is represented by $\xi^0=\pm \infty$
in the coordinate $(\xi^0,{\bf x})$, and so called future and past boundary, respectively.

To further investigate global geometric properties of dS space, we would like to 
introduce other sets of coordinates. \newline
A global coordinate can be introduced by relations
 $$
 \begin{array}{l}
 \xi^0 = \sinh {t}~,         \\
 \xi^i = \cosh {t} \cdot u^{i}~, ~~~( i=1,2,\cdots,n+1 )~,
 \end{array}
 $$
where $u^{i}$ are coordinates of $n$-sphere. It is easy to see that dS
space is topologically equivalent to $ {\bf R} \times  {\bf S}^{n} $.
With this global coordinate, the metric deduced from the $SO(1,n+1)$ invariant one  
is of the form
 \begin{eqnarray}
 ds^2 & = & d\xi^0d\xi^0 - \sum_{i=1}^{n}d\xi^{i}d\xi^{i} - d\xi^{n+1}d\xi^{n+1}  \\
      & = & dt^2 - \cosh^2{t}d\Omega^2_{(n)}~~.  \nonumber
 \end{eqnarray}
Generally, the metric on $n$-sphere can be written as
 $$
 d\Omega^2_{(n)} = dr^2 + \sin^2{r}d\Omega^2_{(n-1)} ~~,
 $$
with $u^{n+1}=\cos{r}$ and $0 \leq r \leq \pi$~, thus, we have
 $$
 ds^2 = dt^2 - \cosh^2{t}dr^2 - \cosh^2{t}\sin^2{r}d\Omega^2_{(n-1)} ~~.
 $$
Suppressing the dimension of ${\bf S}^{n-1}$ and using the  diffeomorphism transformation
 $$\left\{
 \begin{array}{ll}
 \eta = 2\tan^{-1}{(e^{t})} ~,~~~~~& (0 < \eta < \pi)~,    \\
 \theta = r                 ~~~~~& (0 \leq \theta \leq \pi)~,
 \end{array}
 \right.
 $$
we obtain
 \begin{eqnarray}
 \displaystyle ds^2 & = & f(\eta)(d\eta^2 - d\theta^2)~,  \\
 \displaystyle f(\eta)& =&  \frac{1}{\sin^2{\eta}} = \cosh^2{t} ~.  \nonumber
 \end{eqnarray}
The Penrose diagram of dS space is shown in figure 1.
%******************************figure 1 **************
\begin{figure}
\epsfxsize=80mm
\epsfysize=80mm
\centerline{\epsffile{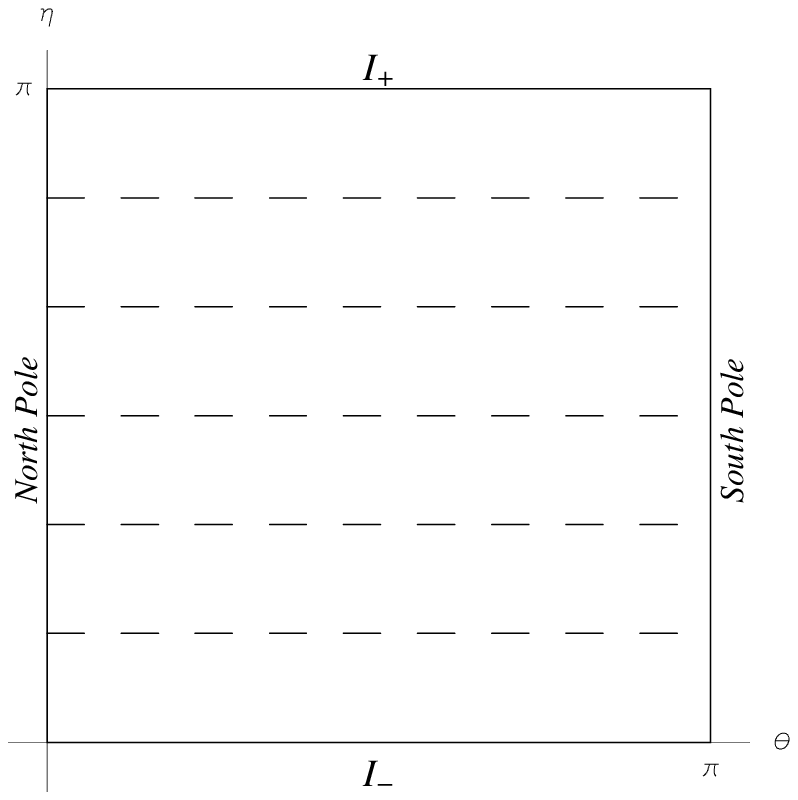}}
\vspace*{0.3cm}
\centerline{ Fig.1~~ Penrose diagram for de-Sitter space }
\end{figure}
%*****************************************************
\newline
Let
 $$\left\{
 \begin{array}{ll}
 \xi^0 + \xi^{n+1} = e^{t}~,               &~~~~  (>0)~,              \\
 \xi^{i} = e^{t} \cdot x^{i} ~,            &~~~~  (i=1,2,\cdots,n)~,  \\
 \xi^0 - \xi^{n+1} = r^2 e^{t} - e^{-t}~,  &~~~~  (r^2 = \sum{x^{i}x^{i}}) ~,
 \end{array}
 \right.
 $$
the metric can be written as

 \begin{equation}
 ds^2  =  dt^2 - e^{2t}\sum_{i=1}^{n}dx^{i}dx^{i} ~~,
 \end{equation}
where the subspace of $(x^{i})$ or $(t={\rm constant})$ is Euclidean.
 \newline
With relation

$$
\displaystyle e^{t} = \xi^0+\xi^{n+1}=\frac{\cos{\theta} - \cos{\eta}}{\sin{\eta}} > 0 ~~,
$$
we can see clearly that this coordinate describes upper-left half of dS space as shown 
in figure 2.
%**************************figure 2******************
\begin{figure}
\epsfxsize=80mm
\epsfysize=80mm
\centerline{\epsffile{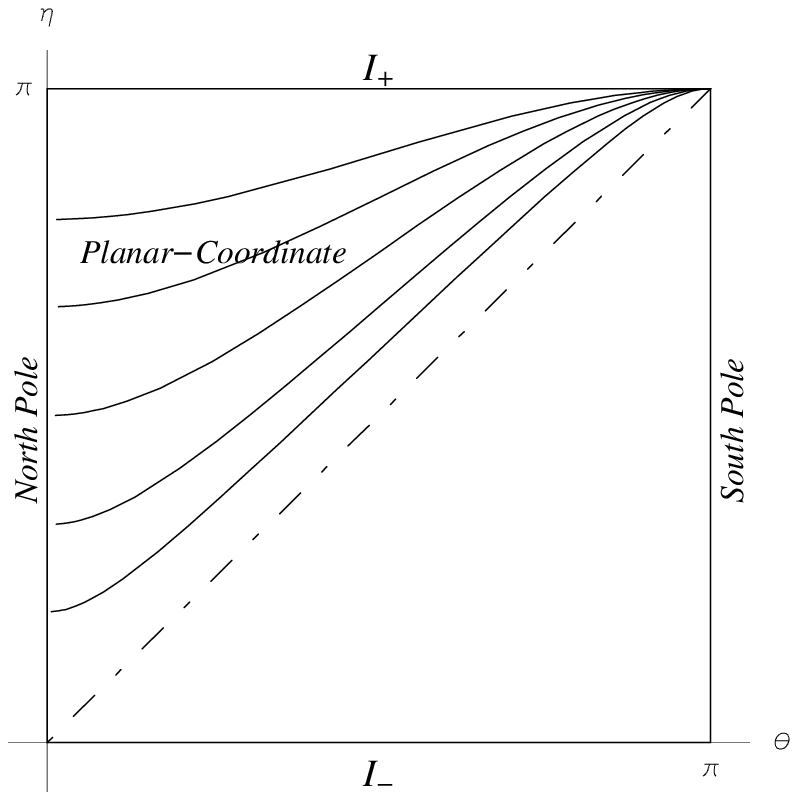}}
\vspace*{0.3cm}
\centerline{Fig.2~~ Planar coordinate for de-Sitter space }
\end{figure}
%****************************************************
\newline
Let

 $$\left\{
 \begin{array}{lll}
 \xi^0 + \xi^{n+1}    & = (1-r^2)^{1/2}e^{t}~,   &~~  (>0)~,    \nonumber  \\
 \xi^0 - \xi^{n+1}    & = -(1-r^2)^{1/2}e^{-t}~, &~~  (<0)~,    \nonumber  \\
 \sum{\xi^{i}\xi^{i}} & = r^2 ~,                 &~~  (<1)~,~(i=1,2,\cdots,n),  \nonumber \\
 \end{array}
 \right.
 $$
then, we get

 \begin{equation}
 \displaystyle ds^2 = (1-r^2)dt^2 - \frac{dr^2}{1-r^2} - r^2d\Omega^2_{(n-1)} ~~.
 \end{equation}
 \newline
With relations

 $$\left\{
 \begin{array}{l}
 \displaystyle \xi^0 + \xi^{n+1} = \frac{\cos{\theta} - \cos{\eta}}{\sin{\eta}} > 0 ~,\\
 \displaystyle \xi^0 - \xi^{n+1} = -\frac{\cos{\theta} + \cos{\eta}}{\sin{\eta}}< 0 ~,
 \end{array}
 \right.
 $$
we can see that this coordinate describes a quarter of dS space as shown in figure 3.
%***********************figure 3**************
\begin{figure}
\epsfxsize=80mm
\epsfysize=80mm
\centerline{\epsffile{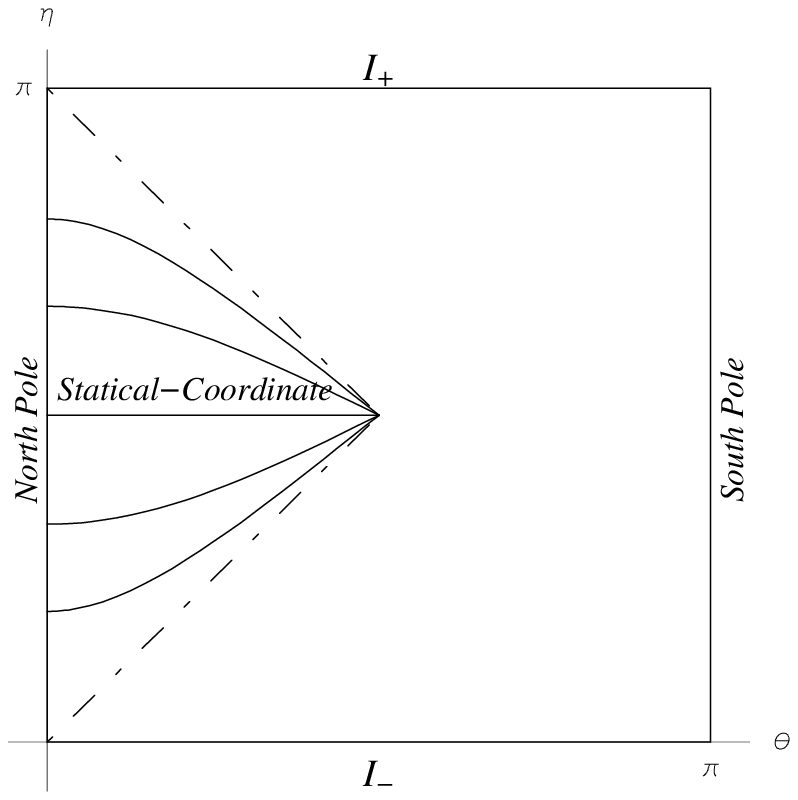}}
\vspace*{0.3cm}
\centerline{ Fig.3~~ Statical coordinate for a slice of de-Sitter space }
\end{figure}
%*********************************************
\newline
\section{Dynamics of massive scalar fields}

The equation of motion for massive scalar fields is of the form

 \begin{equation}\label{KG}
 (\Box + m_0^2) \phi (x) = 0 ~~,
 \end{equation}
where $\Box$ denotes the Laplace operator  in dS space
 \begin{eqnarray}
 \displaystyle
 \Box &=&\frac{1}{\sqrt{g}}\partial_{i}(\sqrt{g}g^{ij}\partial_{j}) \nonumber \\
      &=&(1+\xi^0\xi^0)\partial_{\xi^0}^2+(n+1)\xi^0\partial_{\xi^0}        \nonumber \\
      & &-(1+\xi^0\xi^0)^{-1}\left[(1+\rho^2)^2\partial_{\rho}^2+\rho^{-1}(1+\rho^2)
          (n-1+2\rho^2)\partial_{\rho} \right]                                        \\
      & &-(1+\xi^0\xi^0)^{-1}(1+\rho^2)\rho^{-2}\Delta_{(n-1)} ~~,          \nonumber
 \end{eqnarray}
here $\Delta_{(n-1)}$ is the Laplace operator on ${\bf S}^{n-1}$. \newline 
Rewriting the scalar field $\phi$ into variable-separating form
 $$
 \phi(\xi^0,\rho,{\bf \Theta})={\cal T}(\xi^0){\cal R}(\rho)Y_{l{\bf m}}({\bf \Theta})~,
 $$
we can transform the equation (\ref {KG}) as
 \begin{eqnarray}
  && (1+\xi^0\xi^0)^2{\cal T}''(\xi^0)+(n+1)\xi^0(1+\xi^0\xi^0){\cal T}'(\xi^0)
     +[m_0^2(1+\xi^0\xi^0)+(\epsilon-m_0^2)]{\cal T}(\xi^0)=0  ~,         \nonumber \\
  && \rho^2(1+\rho^2){\cal R}''(\rho)+(n-1+2\rho^2)\rho{\cal R}'(\rho)
     +[\rho^2(1+\rho^2)^{-1}(\epsilon-m_0^2)-l(l+n-2)]{\cal R}(\rho)=0 ~, \nonumber \\
  && [\Delta_{(n-1)} + l(l+n-2)]Y_{l{\bf m}}({\bf \Theta})=0 ~~,
 \end{eqnarray}
where $Y_{l{\bf m}}({\bf {\Theta}})$ is the spherical harmonic function on
${\bf S}^{n-1}$ and $\epsilon$ an arbitrary constant.
 \newline
Solutions for the time and radial parts are, respectively,

 \begin{eqnarray}
 &{\cal T}(\xi^0)&=(1+\xi^0\xi^0)^{(1-n)/4}\cdot\left\{
                   \begin{array}{l}
                   P_{\nu}^{\mu}(i\xi^0) \\
                   \newline              \\
                   Q_{\nu}^{\mu}(i\xi^0)
                   \end{array}\right. ~,   \\
                   \newline \nonumber    \\
 &{\cal R}(\rho)&=\rho^l(1+\rho^2)^{k/2}F\left(\frac{1}{2}(l+k+1),\frac{1}{2}(l+k),
  l+\frac{n}{2};-\rho^2\right) ~~,
 \end{eqnarray}
where we have used the notations

  \begin {equation} \label{notation}
  \left\{
  \begin{array}{l}
  \displaystyle \mu^2 = \frac{1}{4}(n-1)^2 + (\epsilon-m_0^2)~, \\
  \displaystyle \nu(\nu+1) = \frac{1}{4}(n^2-1) - m_0^2  ~,     \\
  \displaystyle k^2 - (n-1)k - (\epsilon-m_0^2) = 0  ~.
  \end{array}
  \right.
  \end {equation}
\newline
Now we consider properties of ${\cal T}(\xi^0)$ near the boundary
$\xi^0\xi^0=\infty$.
\newline
For the first kind of associative Legendre function $P_{\nu}^{\mu}(i\xi^0)$, 
by making use of the formula

 \begin{eqnarray}
 \displaystyle
 P_{\nu}^{\mu}(z) = &&\frac{2^{\nu}\Gamma(\nu+\frac{1}{2})z^{\nu+\mu}
                      (z^2-1)^{-\mu/2}}{\Gamma(\frac{1}{2})\Gamma(1+\nu-\mu)} \\       
                    &&\times F\left(\frac{1-\nu-\mu}{2}, -\frac{\nu+\mu}{2},
                      \frac{1}{2}-\nu, z^{-2} \right)                         \nonumber\\
                    &&+\frac{2^{-\nu-1}\Gamma(-\nu-\frac{1}{2})z^{-\nu+\mu-1}
                      (z^2-1)^{-\mu/2}}{\Gamma(\frac{1}{2})\Gamma(-\nu-\mu)}  \nonumber\\
                    &&\times F\left(\frac{2+\nu-\mu}{2}, \frac{1+\nu-\mu}{2},
                      \frac{3}{2}+\nu, z^{-2} \right) ~~,                     \nonumber
 \end{eqnarray}
we know that  $P_{\nu}^{\mu}(i\xi^0)$ is divergent  near boundary as

 \begin{equation}
 P_{\nu}^{\mu}(i\xi^0) \sim (\xi^0)^{\nu}+(\xi^0)^{-\nu-1}~.
\end{equation}
For the second kind of associative Legendre function $Q_{\nu}^{\mu}(i\xi^0)$,
by making use of the formula

 \begin{eqnarray}
 \displaystyle
 Q_{\nu}^{\mu}(z) = &&\frac{e^{\mu\pi i}}{2^{\nu+1}}\frac{\Gamma(\nu+\mu+1)
                       \Gamma(\frac{1}{2})}{\Gamma(\nu+\frac{3}{2})}(z^2-1)^{\mu/2}
                       z^{-\nu-\mu-1}          \nonumber \\
                    &&\times F\left( \frac{\nu+\mu+1}{2}, \frac{\nu+\mu+2}{2},
                       \nu+\frac{3}{2}, z^{-2} \right) ~,
 \end{eqnarray}
we know  that $Q_{\nu}^{\mu}(i\xi^0)$ is divergent near boundary as

 \begin{equation}
 Q_{\nu}^{\mu}(i\xi^0) \sim (\xi^0)^{-\nu-1} ~.
 \end{equation}

\section{The dS/CFT correspondence}

One of the keystones of the dS/CFT duality is the bulk-boundary propagator. 
To discuss the dS/CFT correspondence explicitly, we have to construct a bulk-boundary 
propagator for dS space. 
Here, by making use of the exact solutions of the equations of motion for massive
scalar fields obtained in the last section, we write down a bulk-boundary propagator,\\

 $G_{B\partial}^\pm(\xi^0,\rho,{\bf u};\varrho,{\bf v})$
 \begin{eqnarray}
 &=&\displaystyle\int d\epsilon\sum_l\sum_{\bf m}Y_{l{\bf m}}({\bf u}-{\bf v})
    (1+\xi^0\xi^0)^{(1-n)/4}(\rho-\varrho)^l(1+(\rho-\varrho)^2)^{k/2}   \\
 && \displaystyle\times P^{\mu}_{\nu}(i\xi^0) \cdot 
    F\left(\frac{1}{2}(l+k+1),\frac{1}{2}(l+k),l+\frac{n}{2};-(\rho-\varrho)^2\right)~,\nonumber
 \end{eqnarray}
where $\mu$, $\nu$ and $k$ take the same values as in the equation (\ref{notation}).
It is not difficult to see that this bulk-boundary propagator satisfies the equations 
of motion for a scalar field with mass $m_0$,

 \begin{equation}
 \left(~\Box+m^2_0~\right)G_{B\partial}^\pm(\xi^0,\rho,{\bf u};\varrho,{\bf v})=0~.
 \end{equation}
 \newline
The bulk field can be determined from a field living on the boundary by making use of 
the bulk-boundary propagator $G^\pm_{B\partial}(\xi^0,\rho,{\bf u};\varrho,{\bf v})$,

 \begin{equation}\label{formula}
 \Phi^\pm(\xi^0,\rho,{\bf u})=\frac{1}{\omega_{n}}\int d{\bf v}
 d\varrho G^\pm_{B\partial}(\xi^0,\rho,{\bf u};\varrho,{\bf v})\phi(\varrho,{\bf v})~,
 \end{equation}
where $\omega_n$ denotes the normalization constant. This shows that
$G^\pm_{B\partial}(\xi^0,\rho,{\bf u};\varrho,{\bf v})$ is really a bulk-boundary propagator 
in dS space for massive scalar fields.  The bulk field $\Phi(\xi^0,\rho,{\bf u})$ 
got from this way does not approach to $\phi(\varrho,{\bf v})$ 
when limited on the boundary, but goes divergent at boundary with dimension

 \begin{equation}
 d=\frac{n}{2}\left(1\mp\sqrt{1-\frac{4m_0^2}{n^2}}\right)~.
 \end{equation}
This is in agreement with the mass bound given by Strominger.

\vskip 2mm

\centerline{\bf Acknowledgments}

We would like to thank H. Y. Guo and C. G. Huang  for enlightening discussions.
The work was supported in part by the National Natural Science Foundation of China.

\end{document}